\documentclass[dvipsnames]{article}
\usepackage{fullpage}
\usepackage[square,numbers]{natbib}
\usepackage{color}
\usepackage{amsmath}
\usepackage{xcolor}
\usepackage{hyperref}
\usepackage[inline, shortlabels]{enumitem}
\usepackage{algorithm}
\usepackage{algpseudocode} 
\usepackage{graphicx}
\usepackage{subcaption}

\setlist{itemjoin ={,\enspace},itemjoin* = { and\enspace}}

\newcommand{\frSet}[1]{\mathcal{#1}}

\newcommand{\frDocument}[0]{d}
\newcommand{\frDocuments}[0]{\frSet{D}}

\newcommand{\frQueries}[0]{\frSet{Q}}
\newcommand{\frQuery}[0]{q}
\newcommand{\frAuthor}[0]{a}
\newcommand{\frAuthors}[0]{\frSet{A}}
\newcommand{\frExposure}[0]{e}
\newcommand{\frGroupExposure}[0]{\mathcal{E}}

\newcommand{\frRanking}[0]{\pi}
\newcommand{\frRankings}[0]{\Pi}

\newcommand{\frRelevance}[0]{r}
\newcommand{\frRankingEffectiveness}[0]{u}
\newcommand{\frEffectiveness}[0]{\text{U}}
\newcommand{\frGroupRelevance}[0]{\mathcal{R}}
\newcommand{\frGroups}[0]{\frSet{G}}
\newcommand{\frGroup}[0]{g}
\newcommand{\frNumGroups}[0]{|\frGroups|}

\newcommand{\frRelevanceTransform}[0]{f}

\newcommand{\frNormalizedExposure}[0]{\Delta}
\begin{document}
\title{Overview of the TREC 2019 Fair Ranking Track\footnote{Data and code are available at: \url{https://fair-trec.github.io/2019/}}}
\author{
  Asia J. Biega\\
  Microsoft Research Montr\'eal\\
  \texttt{asia.biega@microsoft.com}\\
  \\
  Michael D. Ekstrand\\
  Boise State University\\
  \texttt{michaelekstrand@boisestate.edu}
  \and
  Fernando Diaz\\
  Microsoft Research Montr\'eal\\
  \texttt{diazf@acm.org}\\
  \\
  Sebastian Kohlmeier\\
  Allen Institute for Artificial Intelligence\\
  \texttt{sebastiank@allenai.org}
}
\date{\empty}
\maketitle

\begin{abstract}
    The goal of the TREC Fair Ranking track was to develop a benchmark for evaluating retrieval systems in terms of fairness to different content providers in addition to classic notions of relevance. As part of the benchmark, we defined standardized fairness metrics with evaluation protocols and released a dataset for the fair ranking problem. The 2019 task focused on reranking academic paper abstracts given a query. The objective was to fairly represent relevant authors from several groups that were unknown at the system submission time. Thus, the track emphasized the development of systems which have robust performance across a variety of group definitions. Participants were provided with querylog data (queries, documents, and relevance) from Semantic Scholar. This paper presents an overview of the track, including the task definition, descriptions of the data and the annotation process, as well as a comparison of the performance of submitted systems.

\end{abstract}

\section{Introduction}
Modern information access systems influence both the \emph{consumers} and the \emph{producers} of the content that they mediate.
Some systems are quite explicitly two-sided, such as online dating platforms \cite{diaz:sigir2010}.
Hiring platforms, where employers ``consume'' rankings of job-seekers, 
and micro-lending platforms where recommender algorithms help lenders identify people they wish to invest in \cite{burke:fatml2017}, are also
clearly two-sided.
More traditional environments, such as
music, video, and book recommendation systems where direct users consume rankings of works by different artists;
community question answering forums where some members consume answers written by other members;
and various e-commerce platforms where customers consume rankings of products offered by different sellers
can also be considered two-sided because of the indirect matching that happens between the users and content creators.
To date, however, evaluation methodologies in information retrieval and recommender systems have largely focused on quantifying the experience of the consumers through metrics focused on accuracy, diversity, and novelty.

Failing to adequately measure and address the effect of information access on content producers can have significant impact, both socially and
to the information platform.
If content producers feel that a platform does not enable them to connect to their audience effectively, they may move their content to a different platform (harming the platform's inventory) or leave the industry entirely.
A system in which different groups of content producers experience disparate discoverability will harm the less-discoverable groups; if that division falls along lines of historical inequities, such as a music service where indigenous musicians have more difficulty being discovered, it may reinforce or even exacerbate biases and discrimination.
Ineffective promotion of less popular musical genres may also distort the development of culture more broadly \cite{mehrotra:cik2018}. Similar problems were recognized by the early work on fair rankings~\cite{biega:sigir2018, celis2017ranking, singh:kdd2018,wu2018discrimination, yang:ssdb2017, zehlike:cikm2017}.

\noindent The goals of the Fair Ranking track are to,
\begin{itemize}
    \item develop metrics for fair exposure of individuals or groups in retrieval scenarios,
    \item design an experimentation protocol for fair ranking,
    \item release a data set for benchmarking fair ranking algorithms, and
    \item promoting the development of fair ranking algorithms.
\end{itemize}

For 2019, we adopted an \textbf{academic search task}, where we have a corpus of academic article abstracts and queries submitted to a production academic search engine.   The central goal of the Fair Ranking track is to provide \textbf{fair exposure to different groups of authors} (a \emph{group fairness} framing).  

We recognize that there may be multiple group definitions (e.g. based on demographics, stature, topic) and hoped for the systems to be robust to these.  As such, participants were expected to develop systems to optimize for fairness and relevance for \textbf{arbitrary group definitions}, and we did not reveal the exact group definitions until \emph{after} the evaluation runs were submitted. 

The track was set up as a \textbf{reranking} task. We provided participants with a sequence of queries, each accompanied by a varying-size set of documents for each query;
the task was to rerank the documents to produce result lists that are fair and relevant.

\section{Task Description}
Prior work in the area of fair ranking shows that reasoning about fairness should be done over sequences of rankings rather than individual rankings~\cite{biega:sigir2018,singh:kdd2018}.
For the evaluation in this track, we thus provided participants with a sequence $\frQueries$ of queries accompanied by unordered sets of documents to rank.
The document sets were of varying size. For each request (query $\frQuery$ and set of documents $\frDocuments_\frQuery$), participants provided a ranked list of the documents from $\frDocuments_\frQuery$. The final system output was a sequence of rankings. Algorithm~\ref{protocol} presents a pseudocode of the evaluation protocol. 

Participants were instructed to make their system optimize the ranking sequence for two goals:
\begin {enumerate*} [(1) ]%
\item be relevant to the consumers
\item be fair to the producers.
\end {enumerate*}

\begin{algorithm}
\caption{Evaluation protocol}\label{protocol}
\begin{algorithmic}
\State $\Pi\gets \{\}$
\For{$\frQuery,\frDocuments_\frQuery\in\frQueries$}
\State $\pi\gets \Call{\textcolor{red}{System}}{\frQuery, \frDocuments_\frQuery}$
\State $\Pi\gets \Pi + [\pi]$
\EndFor
\State \Return $\Pi$
\end{algorithmic}
\end{algorithm}

\section{Evaluation}
\label{sec:evaluation}
Unlike previous TREC tracks, participants received multiple copies of the same query text--with varying query ids--and were allowed to submit different rankings for each instance of the query.  At evaluation time, we measured \emph{amortized performance} over rankings produced for each given query, as well as across all rankings and queries (macro- and micro- amortization, respectively.)  

Given a sequence of queries $\frQueries$ and associated system rankings, we evaluate systems according to fair exposure of authors (Section \ref{sec:attention}) and relevance of documents (Section \ref{sec:relevance}).  

\subsection{Measuring Fairness}
\subsubsection{Measuring Author Exposure for a Single Ranking}
\label{sec:attention}
\begin{figure}
	\begin{center}
	\includegraphics[width=2in]{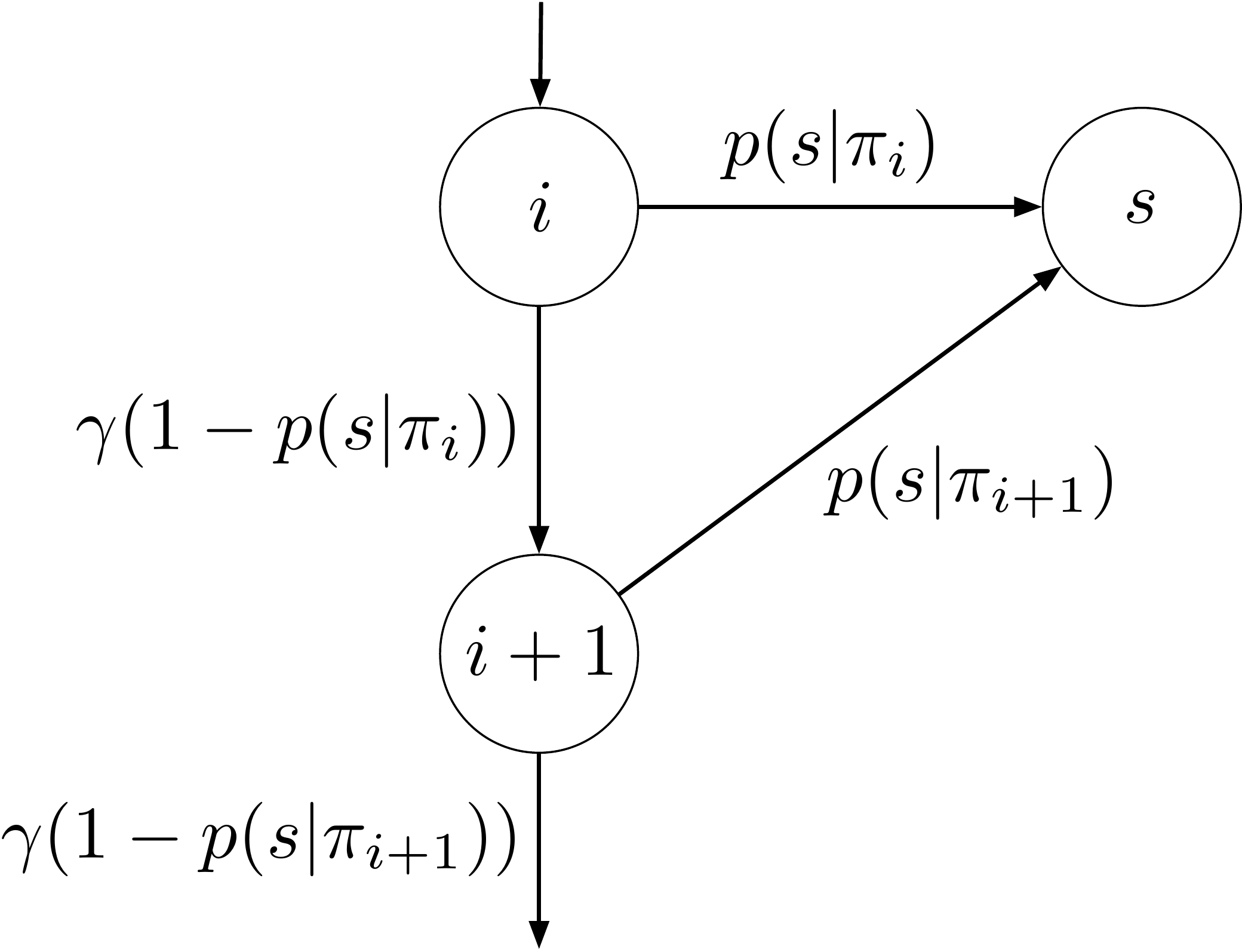}
	\end{center}
	\caption{Attention model.}
	\label{fig:attention}
\end{figure}

In order to measure exposure, we adopt the browsing model underlying the Expected Reciprocal Rank metric \cite{chapelle:err}.  Given a ranking $\frRanking$, the exposure of author $\frAuthor$ is,
\begin{align}
\begin{split}
\frExposure_\frAuthor^\frRanking &= \sum_{i=1}^n \left[\gamma^{i-1}\prod_{j=1}^{i-1}(1-p(s|\pi_j))\right] I(\pi_i\in\frDocuments_\frAuthor)\\
\\
n & \phantom{=} \text{number of documents in ranking }\pi\\
\frDocuments_\frAuthor & \phantom{=} \text{documents including $\frAuthor$ as an author}\\
\pi_i & \phantom{=} \text{document at position $i$}\\
\gamma & \phantom{=} \text{continuation probability (fixed to 0 for the final position in the ranking)}\\
p(s|\frDocument) & \phantom{=} \text{probability of stopping given user examined $\frDocument$}
\end{split}
\label{fig:exposure}
\end{align}
We present a graphical depiction of this model in Figure \ref{fig:attention}.

We fixed the value of the discounting factor $\gamma$,
and assumed $p(s|\frDocument) = \frRelevanceTransform(\frRelevance_\frDocument)$, where $\frRelevance_\frDocument$ is the relevance of the document $\frDocument$ and $\frRelevanceTransform$ is a monotonic transform of that relevance into a probability of being satisfied.

We compute the amortized exposure for $\frAuthor$ as,
\begin{align}
	\frExposure_\frAuthor &= \sum_{\frRanking\in\frRankings} \frExposure_\frAuthor^\frRanking
\end{align}
where $\frRankings$ is the sequence of all system rankings.  
\subsubsection{Measuring Author Relevance for a Single Ranking}
The author relevance for a ranking $\frRanking$ is defined as,
\begin{align}
	\frRelevance_\frAuthor^\frRanking &= \sum_{\frDocument \in \frDocuments_\frAuthor}p(s|\frDocument)
	\label{fig:relevance}
\end{align}
Notice that this metric is independent of the ranking but not the query.  As with amortized exposure, we define amortized relevance as the sum over all rankings $\frRankings$.  

\subsubsection{Measuring Group Fairness}
Assume that each author is assigned to exactly one of $\frNumGroups$ groups.  Let $\frAuthors_\frGroup$ be the set of all authors in group $\frGroup$.  The group exposure and relevance metrics are defined as,
\begin{align}
	\frGroupExposure_\frGroup &= \frac{\sum_{\frAuthor \in \frAuthors_\frGroup} \frExposure_\frAuthor}{\sum_{\frGroup'\in\frGroups}\sum_{\frAuthor \in \frAuthors_{\frGroup'}} \frExposure_\frAuthor}\\
	\frGroupRelevance_\frGroup &= \frac{\sum_{\frAuthor \in \frAuthors_\frGroup} \frRelevance_\frAuthor}{\sum_{\frGroup'\in\frGroups}\sum_{\frAuthor \in \frAuthors_{\frGroup'}} \frRelevance_\frAuthor}
\end{align}

We assume that groups should receive exposure proportional to relevance.  We adopt the following measure to quantify the deviation from this ideal exposure,
\begin{align}
	\frNormalizedExposure_{\frGroup} &=  \frGroupExposure_\frGroup - \frGroupRelevance_\frGroup 
\end{align}

Given this relevance-normalized measure of exposure, we can compute the fair exposure using the $\ell_2$ norm,
\begin{align}
	\frNormalizedExposure &= \sqrt{\sum_{\frGroup\in\frGroups} \frNormalizedExposure_{\frGroup}^2}
\label{eq:unfairness}
\end{align}
Because this metric does not capture some of the nuance of how over- and under-exposure is distributed, we will adopt secondary metrics in our final analysis of results.

\subsection{Measuring Relevance}
\label{sec:relevance}
We measured the quality of a ranking for the searchers as the expected utility, assuming the same attention model as used for our fairness metric,
\begin{align}
\frRankingEffectiveness^\frRanking &= \sum_{i=1}^{n}\left[\gamma^{i-1}\prod_{j=1}^{i-1}(1-p(s|\pi_j))\right]p(s|\pi_i)
\label{eq:utility}
\end{align}
We average all utilities of rankings, $\frEffectiveness=\frac{1}{|\frRankings|}\sum_{\frRanking \in \Pi} \frRankingEffectiveness^\frRanking$ as our final relevance metric.

\subsection{Trading Off Fairness and Relevance}
Although a system could, in theory, achieve the optimal relevance and fairness, in practice, relevance will degrade as fairness improves.  We therefoe measured the trade-offs between fairness to producers and quality for consumers as an auxiliary metric.

\section{Data}

\label{sec:data}

\subsection{Input}

There were three main inputs available to participants: the \emph{corpus} of articles to rank, the \emph{example group definition} file
to help develop and test solutions, and the \emph{queries}.

\subsubsection{Corpus}

The corpus for this track was the Semantic Scholar (S2) Open Corpus from the Allen Institute for Artificial Intelligence.
It can be downloaded from \url{http://api.semanticscholar.org/corpus/}, and consists of 47 1GB data files. Each file is
compressed JSON, where each line is a JSON object describing one paper. 
\newpage 
\noindent  The following data are available for most papers:

\begin{itemize}
\itemsep-0.1em 
    \item S2 Paper ID
    \item DOI
    \item Title
    \item Abstract
    \item Authors (resolved to author IDs)
    \item Inbound and outbound citations (resolved to S2 paper IDs)
\end{itemize}

\subsubsection{Queries}
\label{sec:queries}
\paragraph{Query data.} Query sequences were constructed based on a query log provided by Semantic Scholar\footnote{\url{https://www.semanticscholar.org}}. 
This query log specified, for each query, a list of documents that were displayed at each ranking position (up to 6), and the number of clicks observed for each document. In case the list of documents displayed in the top positions changed over time for a query, we merged all the documents observed as a response to the query into the query reranking pool. 

Because of the exhaustive annotation process that required annotating group memberships of all document authors, we sampled a smaller number of queries to construct evaluation sequences. We released $652$ training and $635$ evaluation queries. For both the training and evaluation data, these queries were selected first by random sampling, and then by a number of filtering steps. More specifically,
\begin{itemize}
    \item We included only queries with at least three observed clicks, and where the clicks occur at at least two different ranking positions (this heuristic helped remove noise and filter out many of the known-item queries).
    \item We further manually cleaned the sample to remove any known-item queries, queries containing people's names, and queries with offensive and sensitive keywords.
    \item For evaluation queries, we additionally considered whether the documents in reranking pools appeared in the S2 corpus. We kept queries that had at least 5 resolvable documents, at least 1 of which has been clicked.
\end{itemize}
Data that was made publicly available contained, for each query:

\begin{itemize}
\itemsep-0.1em 
    \item Query ID,
    \item Query string,
    \item Query normalized frequency,
    \item A list of documents to rerank (with relevance for training queries, and without relevance for evaluation queries).
\end{itemize}
Relevance was binary. We considered a document to be relevant to a query if there was at least one observed click on that document for the query, and not relevant otherwise.
\paragraph{Query sequences.} Runs were submitted over \emph{query sequences}: ordered sequences of queries that may contain duplicates. For training and development, we provided participants with training queries (with relevance and frequency data) and a script to generate query sequences. For evaluation, we provided query data (without relevance) not included in the training data, and generated five evaluation query sequences, each containing $25$k queries. Sequences were generated by sampling queries from a distribution based on query frequencies.

\subsubsection{Annotations}
NIST assessors annotated returned papers with the country in which each author was operating (based on their affiliation data in the paper manuscript). 
Not all papers were able to be annotated.  These are the known reasons a paper may not have annotations:

\begin{itemize}
    \item It has a large author list ($> 25$). We excluded such long papers because there were not very many of them, and large-team papers require special treatment in how we consider their author lists, particularly when authors may be from different groups.
    \item Due to a bug in the large-paper exclusion logic, a number of papers with a medium number of authors were also excluded.
    \item Some papers did not have an accessible source with sufficient affiliation information to provide annotations (e.g. no available PDF file, and a paper information page that either did not contain affiliation details or was not accessible from the annotation interface).
    \item Some papers may not provide sufficient information to determine an author's affiliation location.
\end{itemize}

\begin{table}[]
    \centering
    \begin{tabular}{lr}
        Documents & 5,620  \\
        Annotated Documents & 2,866 \\
        Have Country Data & 2,823 \\
        Advanced Econ Papers & 2,372 \\
        Developing Econ Papers & 308 \\
        Mixed Econ Papers & 138 \\
        Advanced Econ Authors & 7,018 \\
        Developing Econ Authors & 1,184
    \end{tabular}
    \caption{Annotation Outcome Summary}
    \label{tab:outcomes}
\end{table}

Table~\ref{tab:outcomes} shows the coverage of annotations, and the number in each group, after merging and integrating data sources.  For these statistics, to aggregate each paper's authors into a single economic designation for the paper, we considered a paper to be from an advanced or developing economy if all authors' locations had the same economic designation; otherwise, we list it as a `mixed' economy paper.

\subsubsection{Group Definitions}
\paragraph{Group definition accompanying the training data.}
To help participants get started, we provided a file 
containing group membership definitions for authors in the S2
corpus. This definition was based on author i--10 indices (i--10 denotes the number of papers with at least 10 citations a person has coauthored).  This definition was not used in the final evaluation, but was meant as a starting point for system development. For each author, the data consisted of:
\begin{itemize}
\itemsep-0.1em 
    \item the author's S2 ID,
    \item the author's group identifier.
\end{itemize}
Authors were split into $7$ groups, based on the value of their i--10 index.

\paragraph{Group definitions for evaluation.}
For evaluation, we used two different statistics to derive group definitions. The first definition was based on the NIST assessors' country annotations. We combined these annotations with economic development levels from the International Monetary Fund. With this definition, the fairness target is to ensure fair exposure for papers written in countries with more- and less-developed economies. The evaluation itself uses individual author-level annotations; the exposure a mixed-economy paper receives counts towards both developing and advanced economy exposure. Under this definition, authors are split into two groups.

The second definition is similar to the training group definition; but based on the h-index (h-index denotes the number $h$ of papers with at least $h$ citations a person has coauthored) of paper authors. Authors are split into four groups based on the value of their h-index: $h < 5$, $5 \leq h < 15$, $15 \leq h < 30$, $h \geq 30$.

\subsection{Output}

Each run output was submitted as a JSON file with the following contents:

\begin{itemize}
    \item qnum: $\langle \text{sequence id}\rangle.\langle \text{query number in sequence}\rangle$
    \item qid (to look up in the query file): $\langle \text{query id}\rangle$
    \item ranking: an ordered list of document IDs (of the documents to be reranked for the query).
\end{itemize}

\section{Results}

\subsection{Evaluation parameters}
The results in this notebook were computed using the following parameter values:
\begin{itemize}
    \item Continuation probability (Eq.~\ref{fig:exposure}): $\gamma=0.5$;
    \item Stopping probability given a document (Eq.~\ref{fig:exposure} and \ref{fig:relevance}): $p(s|d) = f(r_d) = 0.7 * r_d$.
\end{itemize}
Relevance scores $r_d$ were binary, and computed from the click data as described in Sec.~\ref{sec:queries}.

\subsection{Overall results}
\begin{figure}[t!p]
\centering
       \begin{subfigure}{\textwidth}
               \centering
               \begin{tabular}[b]{lcc}
               run & utility & unfairness \\
               \hline
uognleDivAAsp	&	0.5612	&	0.0059	\\
QUARTZ-e0.00500	&	0.6239	&	0.0191	\\
QUARTZ-e0.01000	&	0.6273	&	0.0198	\\
fair\_random	&	0.5476	&	0.0326	\\
QUARTZ-e0.00010	&	0.6241	&	0.0330	\\
QUARTZ-e0.00001	&	0.6247	&	0.0332	\\
QUARTZ-e0.00100	&	0.6228	&	0.0347	\\
QUARTZ-e0.00200	&	0.6230	&	0.0348	\\
uognleDivAJc	&	0.5544	&	0.0352	\\
first	&	0.5507	&	0.0428	\\
uognleSgbrFair	&	0.6151	&	0.0649	\\
uognleSgbrUtil	&	0.6151	&	0.0649	\\
fair\_LambdaMART	&	0.6599	&	0.0741	\\
MacEwanBase	&	0.6194	&	0.0770	\\
uognleMaxUtil	&	0.6741	&	0.0799	\\
               \end{tabular}~~%
\includegraphics[width=0.6\textwidth]{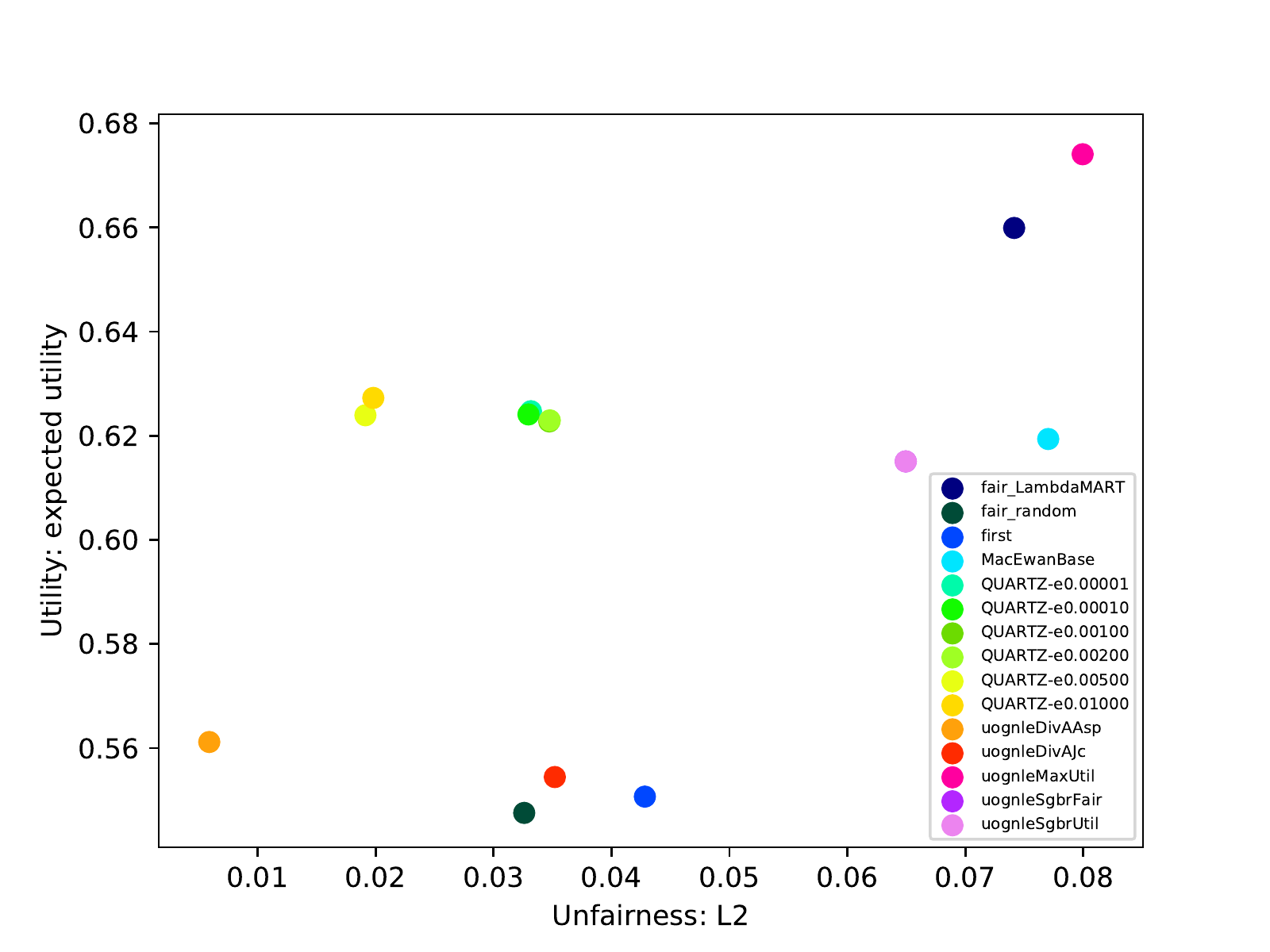}
               \caption{Group definition: IMF level with 2 groups}
               \label{fig:performance-all-level}
     \end{subfigure}
    \begin{subfigure}[b]{\textwidth}
               \centering
                            \begin{tabular}[b]{lcc}
               run & utility & unfairness \\
               \hline
fair\_random	&	0.5476	&	0.0405	\\
uognleDivAJc	&	0.5544	&	0.0449	\\
first	&	0.5507	&	0.0456	\\
MacEwanBase	&	0.6194	&	0.0476	\\
uognleSgbrFair	&	0.6151	&	0.0482	\\
uognleSgbrUtil	&	0.6151	&	0.0482	\\
uognleDivAAsp	&	0.5612	&	0.0585	\\
uognleMaxUtil	&	0.6741	&	0.0656	\\
fair\_LambdaMART	&	0.6599	&	0.0855	\\
QUARTZ-e0.00001	&	0.6247	&	0.1036	\\
QUARTZ-e0.00010	&	0.6241	&	0.1059	\\
QUARTZ-e0.00100	&	0.6228	&	0.1068	\\
QUARTZ-e0.00200	&	0.6230	&	0.1071	\\
QUARTZ-e0.01000	&	0.6273	&	0.1097	\\
QUARTZ-e0.00500	&	0.6239	&	0.1112	
               \end{tabular}~~%
            \includegraphics[width=0.6\textwidth]{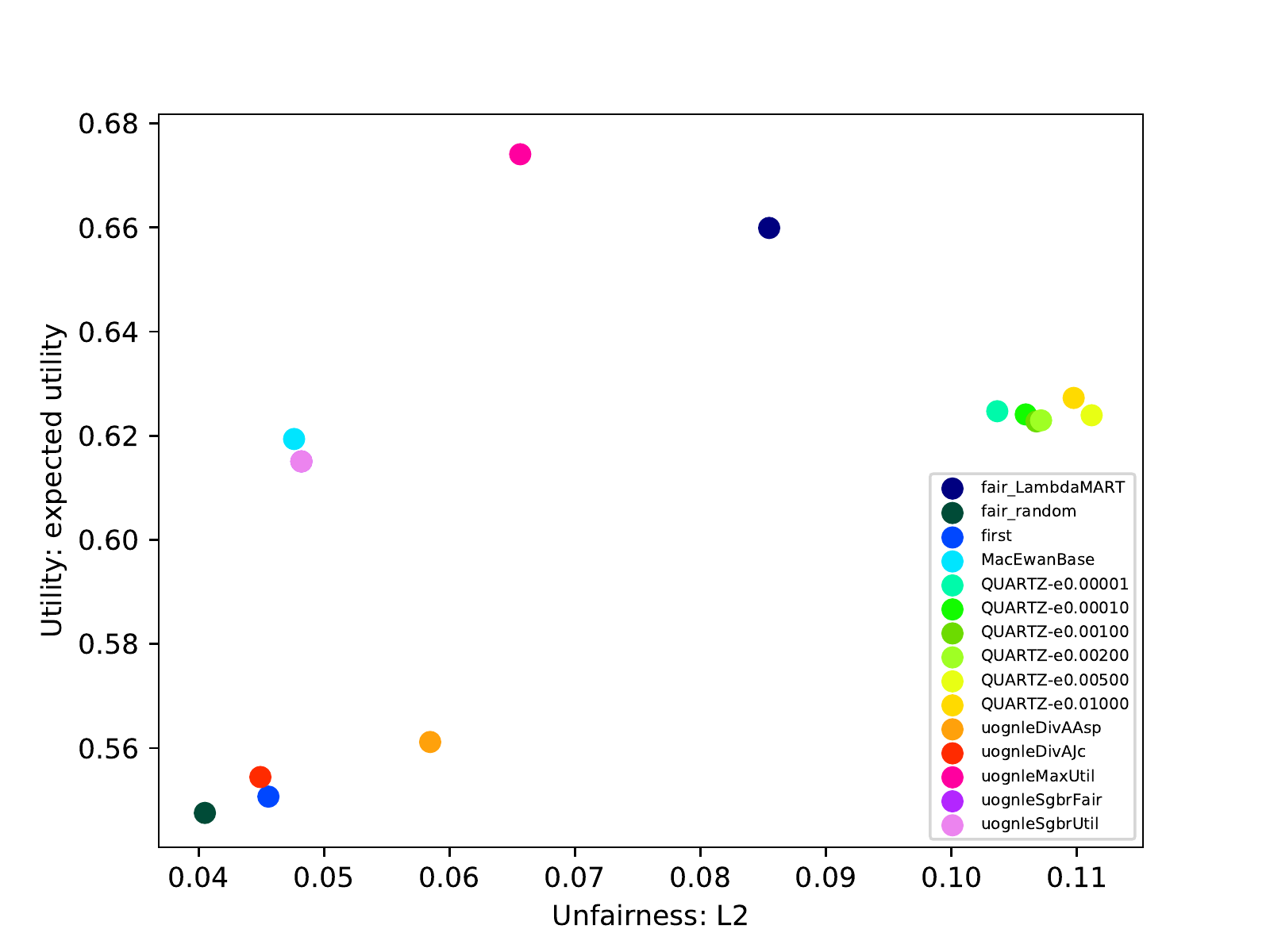}
               \caption{Group definition: h-index with 4 groups}
               \label{fig:performance-all-h4}
     \end{subfigure}
\caption{Performance of all submitted runs in terms of unfairness and expected utility for two different group definitions.  Runs in tables ranked in decreasing order of fairness.  Points toward the upper left area of the figures are preferred to those toward the lower right.}
\label{fig:performance-all}
\end{figure}

Figure~\ref{fig:performance-all} presents the performance of all submitted runs in terms of unfairness (Eq.~\ref{eq:unfairness}) and expected utility (Eq.~\ref{eq:utility}) for two different group definitions (based on the economic levels of author affiliation countries and on h-indices of authors). Note that a run that randomly shuffles the reranked documents (fair\_random) performs well in terms of amortized fairness, but proves to be the worst approach in terms of ranking quality. Based on short run descriptions provided by participants, other runs are based on a number of different approaches: learning to rank without explicit fairness modeling (fair\_LambdaMART), BERT embeddings without explicit fairness modeling (first), an approach where the final ranking is a weighted merge of search results for different fields; weights are adjusted throughout the sequence (MacEwanBase), an approach that models fairness and unfairness distributions over groups (QUARTZ-*), various approaches based on result diversification (uognle*).

We generally observe that the relative ordering of the systems in terms of unfairness is not robust to varying group definitions.

\section{Discussion}

\subsection{Limitations}

There are a number of limitations to the data and evaluation in the current version of the track.
We are actively looking to address several of them in the 2020 edition.

First, we derive a binary relevance construct from click logs.
There are at least three problems with this:

\begin{enumerate}
    \item We do not know if the retrieved document was actually relevant to the user's query, or just looked relevant enough (or surprising enough) to receive a click.
    \item Clicks are going to be biased, possibly with the same biases that we are seeking to correct for.  Our overall fairness goal in this track is to counteract prestige effects, but author recognition and prestige are likely to play a role in determining which article a user will click among articles with comparably relevant titles.
    \item Binary relevance precludes approaches reliant on graded relevance.
\end{enumerate}

Second, our group annotations are quite incomplete.
We do not know the extent to which missingness correlates with protected characteristics, or the impact this has on final system performance.

Third, the rerank task is based on relatively small numbers of documents.
If the source search engine is biased so that the top 6 positions are often given to dominant-group papers, re-ranking systems cannot correct for that bias and give exposure to relevant documents from protected groups that the original search engine ranked at lower positions.

\subsection{TREC 2019 Fair Ranking Track: Lessons Learned}

Running the Fair Ranking track at TREC 2019 taught us a number of valuable lessons that were not apparent in the prior work on fairness in information retrieval,
primarily because published work to date had not had to engage with many of the practicalities of system evaluation.
The primary challenges we faced were related to the fairness metric, label annotations, and data release.

\begin{itemize}
   \item {The fairness metric used in the 2019 task required exhaustive annotation of both group membership of document authors as well as document relevance. Because of this limitation, we decided that the benchmark needs to be a reranking task. It is quite possible that evaluation of this year's results will show that reranking of just a few documents is a trivial fairness problem; we are preparing to revisit the fairness metrics to account both for a recall task and the resource limitations of TREC assessment.}
   \item {Many datasets we considered could not be judged for relevance by non-expert annotators. The dataset of scholarly paper abstracts and queries from Semantic Scholar that the 2019 track was based on contained click information. We used this click information to derive paper relevance. However, our goal in the fairness task was to correct for the tendency of the searchers to select items authored by well-known researchers and institutions. The click information is likely to be biased in exactly this way: Searchers clicking more on abstracts of well-known authors and institutions. This bias can be only quantified after collecting the annotations necessary for producer group assignments.}
   \item {Identifying a sensitive attribute usable in the TREC setting was difficult. Many papers focus on gender, but we cannot use gender for a benchmark when gender identities are not available in existing data, because releasing gender annotations is problematic for ethical and privacy reasons.  Prestige in the university setting is difficult to quantify; while the Carnegie Classification exists for U.S. universities, there is not a mapping other institutions.}
\end{itemize}

\bibliography{fairtrec} 
\end{document}